\documentclass[english,twocolumn]{revtex4-1}
\usepackage[T1]{fontenc}
\usepackage[latin9]{inputenc}
\usepackage{textcomp}
\usepackage{amsmath}
\usepackage{graphicx}

\makeatletter

\newcommand{\lyxmathsym}[1]{\ifmmode\begingroup\def\b@ld{bold}
  \text{\ifx\math@version\b@ld\bfseries\fi#1}\endgroup\else#1\fi}

\makeatother

\usepackage{babel}
\begin{document}

\title{Manipulating Fock states of a harmonic oscillator while preserving
its linearity}

\author{K. Juliusson$^{1}$, S. Bernon$^{1}$, X. Zhou$^{1}$, V. Schmitt$^{1}$,
H. le Sueur$^{2}$, P. Bertet$^{1}$, D. Vion$^{1}$, M. Mirahimi$^{3}$,
P. Rouchon$^{4}$, and D. Esteve$^{1}$}

\affiliation{$^{1}$Quantronics group, SPEC, CEA, CNRS, Université Paris-Saclay,
CEA Saclay, 91191 Gif-sur-Yvette, France.}

\affiliation{$^{2}$Centre de Sciences Nucléaires et de Sciences de la Matière,
91405 Orsay, France.}

\affiliation{$^{3}$INRIA Paris-Rocquencourt, Domaine de Voluceau, B.P. 105, 78153
Le Chesnay Cedex, France.}

\affiliation{$^{4}$Centre Automatique et Systèmes, Mines-ParisTech, PSL Research
University, 60, bd Saint-Michel, 75006 Paris, France.}


\date{\today }
\begin{abstract}
We present a new scheme for controlling the quantum state of a harmonic
oscillator by coupling it to an anharmonic multilevel system (MLS)
with first to second excited state transition frequency on-resonance
with the oscillator{\normalsize{}. In }this scheme that we call 'ef-resonant',{\normalsize{}
the }spurious oscillator Kerr non-linearity inherited from the MLS{\normalsize{}
is very small,} while its Fock states can still be selectively addressed
via an MLS transition at a frequency that depends on the number of
photons. We implement this concept in a circuit-QED setup with a microwave
3D cavity (the oscillator, with frequency 6.4\,GHz and quality factor
$Q_{O}=2\times10^{6}$) embedding a frequency tunable transmon qubit
(the MLS). We characterize the system spectroscopically and demonstrate
selective addressing of Fock states and a Kerr non-linearity below
350\,Hz. At times much longer than the transmon coherence times,
a non-linear cavity response with driving power is also observed and
explained.
\end{abstract}
\maketitle
The harmonic oscillator being an exactly solvable system with a single
degree of freedom, is often taken as a model system in many areas
of physics. In particular, for demonstrating coherent control of simple
quantum systems, physicists have used cold electromagnetic resonators
in the quantum regime, with their quantum states controlled by a single
atom, be it a real one in cavity-QED \cite{Exploring} or an artificial
one in circuit-QED \cite{Blais circuit QED,Wallraff CircuitQED}.
For instance, the preparation of a single photon Fock state by passing
an atom through a cavity \cite{Brune Vaccuum Rabi}, or of an arbitrary
quantum state using the Law and Eberly protocol \cite{LawEberly-1,FockHofheinz,arbitraryHofheinz}
in circuit-QED, are landmark results. In circuit-QED relevant to this
work, the resonator has a frequency $\nu_{O}$ in the microwave domain,
and the artificial atom is a superconducting quantum bit that can
be regarded as an ancillary multilevel system (MLS) with states $\left|g\right\rangle $,$\left|e\right\rangle $,$\left|f\right\rangle $,$\left|h\right\rangle $\dots{}
possibly tunable in energy. The MLS can be used resonantly \cite{FockHofheinz,arbitraryHofheinz},
or off-resonantly \cite{Leghtas Encoding} when its coupling to the
resonator is sufficiently strong to split the transition frequency
$\nu_{ge}$ into different lines at frequencies $\nu_{n}$ that depend
on the photon number $n$ \cite{SchusterResolved}. In this so-called
``photon number split'' regime, any particular Fock state $\left|n\right\rangle $
is addressed at frequency $\nu_{n}$, and symmetrically, the resonator
can be driven coherently at a frequency $\nu_{O,\left|x\right\rangle }$
that depends on the MLS state $\left|x\right\rangle $. This allows
for both manipulating and measuring the oscillator field in various
ways, for instance by MLS-state conditional cavity phase shift \cite{Brune},
photon-number selective phase gates \cite{Heeres - Phase Gates} that
could lead to universal control of the oscillator field \cite{Yale -  Universal control},
or by Wigner and quantum state tomography of the field \cite{Tomo LKB,Kirchmair-TomoP(n)}.

This off-resonant method has been demonstrated with transmon qubits
\cite{Koch-Transmon} coupled dispersively to the resonator, i.e.
with detunings $\left|\varDelta_{xy}=\nu_{xy}-\nu_{O}\right|$ between
any $\nu_{xy}$ transition of the transmon and $\nu_{O}$ much larger
than the coupling frequency $g_{xy}$ between this transition and
the cavity field. In this dispersive regime, the frequencies $\nu_{n}=\nu_{ge}+n\chi$
are equidistant and separated by the so-called dispersive shift $\chi\simeq2\alpha\varepsilon^{2}$
\cite{Koch-Transmon}, with $\alpha=\nu_{ef}-\nu_{ge}$ the transmon
anharmonicity, $\varepsilon=g_{O}/\varDelta$, $g_{O}\equiv g_{ge}$
the transmon-oscillator coupling, and $\varDelta=\nu_{ge}-\nu_{O}$
their detuning. A drawback of this scheme is to transfer part of the
transmon MLS anharmonicity to the oscillator \cite{Black box,Bourasa Limit Kerr},
which can drastically perturb its coherent dynamics \cite{Yin - dynamic kerr,Kirchmair-TomoP(n),YurkeStoler}
and necessitate careful design \cite{Bourasa Limit Kerr} and correction
protocols \cite{Heeres - Phase Gates}. This non-linearity \cite{Koch-Transmon,Black box,Kirchmair-TomoP(n)}
results in a shift of the oscillator frequency $\nu_{O}$ (or Kerr
non-linearity) of about $K=\alpha\varepsilon{}^{4}$ per photon. $K$
scaling as $\chi^{2}/\alpha$ cannot be minimized (at fixed $\alpha$)
without losing the selectivity between Fock states. In this work,
we propose a different transmon-oscillator coupling scheme (see Fig.\ref{fig:resonant_ef})
that yields a much smaller Kerr non-linearity for the same Fock state
selectivity. The idea consists in having the $ef$ transition resonant
with $\nu_{O}$, i.e. $\Delta=-\alpha$, to displace significantly
the qubit excited levels even at small coupling $g_{O}$, while at
the same time staying in the dispersive regime for the first transition
$\nu_{ge}$ to get a small non-linearity. We explain in details in
the next section this 'ef-resonant' condition when the MLS is a slightly
anharmonic three level system, for which analytic results can be obtained.
Then, we describe in section \ref{sec:Experimental-implementation}
our implementation of the proposal, and characterize it experimentally
in section \ref{sec:Experimental-results}.

\section{The 'ef-resonant' coupling scheme for a harmonic oscillator\label{sec:Theoretical-summary}}

\begin{figure}
\includegraphics[width=6cm]{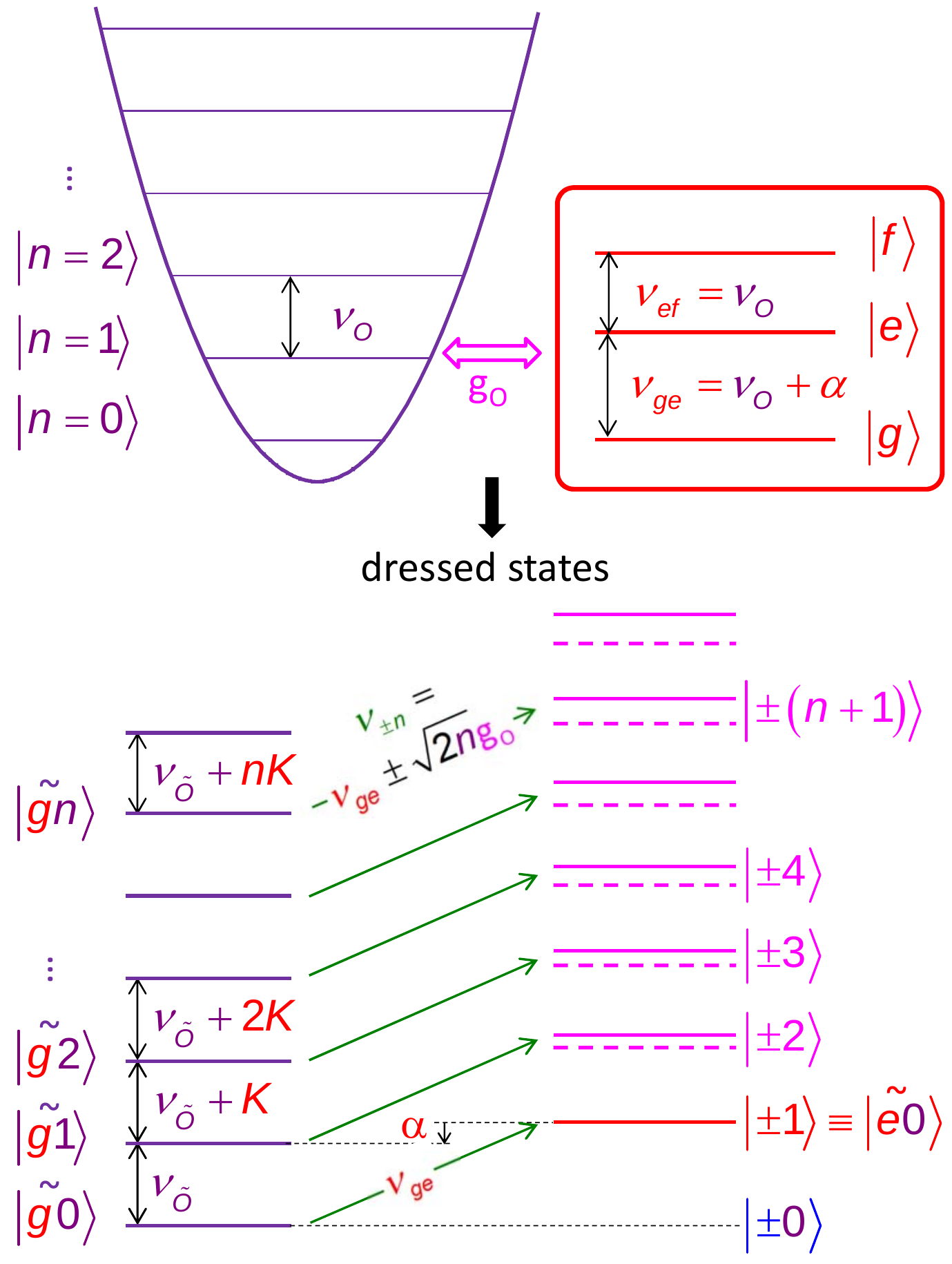} \protect\caption{\label{fig:resonant_ef} 'MLS ef-resonant' scheme for manipulating
Fock states $\left|n\right\rangle $ of a quantum harmonic oscillator.
Left: the oscillator with angular frequency $\nu_{O}$ is coupled
with a coupling frequency $g_{O}$ to a multilevel system (MLS) with
eigenstates$\left\{ \left|g\right\rangle ,\left|e\right\rangle ,\left|f\right\rangle ,...\right\} $,
the $\left|e\right\rangle \leftrightarrow\left|f\right\rangle $ transition
of which is resonant with $\nu_{O}$, whereas the $\left|g\right\rangle \leftrightarrow\left|e\right\rangle $
transition is detuned by $\alpha$. Right: The resulting energy diagram
consists of a quasi-harmonic ladder $\left\{ \left|\widetilde{gn}\right\rangle \right\} $
when the 3LS is left unexcited, and of two anharmonic ladders of levels
$\left|\pm n\right\rangle $ that correspond approximately to symmetric
and anti-symmetric superpositions of $\left|e\left(n-1\right)\right\rangle $
and $\left|f\left(n-2\right)\right\rangle $ states for $n\geq2$.
The $\left|\widetilde{gn}\right\rangle \leftrightarrow\left|\pm\left(n+1\right)\right\rangle $
transitions can be driven at different frequencies $\nu_{ge}\pm\sqrt{2\left(n-1\right)}g_{O}$
to manipulate selectively any $\left|\widetilde{gn}\right\rangle $.}
\end{figure}

To explain the interest of our coupling scheme in the simplest way,
we first consider the case of a harmonic oscillator ($O$) with frequency
$\nu_{O}$ and annihilation operator $a$, coupled to a three-level
system (3LS) with eigenstates $\left\{ \left|g\right\rangle ,\left|e\right\rangle ,\left|f\right\rangle \right\} $
and lowering operator $a_{q}$. We also assume a weak anharmonicity
$\alpha=\nu_{ef}-\nu_{ge}\ll\nu_{ge}$, such that $a_{q}$ can be
approximated by the annihilation operator of a harmonic oscillator
restricted to three levels. The two subsystems obey the resonant condition
$\nu_{ef}=\nu_{O}$ (detuning $\varDelta=\nu_{ge}-\nu_{O}=-\alpha$
) and are subject to an exchange interaction with coupling frequency
$g_{O}\ll\alpha$, yielding the Hamiltonian
\begin{equation}
\begin{array}{l}
H=H_{O}+H_{3LS}+H_{coupling},\\
H_{O}=h\nu_{O}a^{\dagger}a,\\
H_{3LS}=h\left(\nu_{O}a_{q}^{\dagger}a_{q}-\alpha\left|e\right\rangle \left\langle e\right|-\alpha\left|f\right\rangle \left\langle f\right|\right),\\
H_{coupling}=hg_{O}\left(a^{\dagger}a_{q}+aa_{q}^{\dagger}\right).
\end{array}
\end{equation}
This coupling makes the Hamiltonian matrix block-diagonal in the basis
$\left|xn\right\rangle \equiv\left|x\right\rangle \otimes\left|n\right\rangle $
($x=g,e,f$ ), with subsequent blocks of size 1, 2, 3, 3, 3... Using
the reduced coupling $\varepsilon=-g_{O}/\alpha$, these blocks write\begin{widetext}
\begin{equation}
\begin{array}{c}
\left[0\right]_{\left|g0\right\rangle },h\left(\nu_{O}I_{2}-\alpha\left[\begin{array}{cc}
0 & \varepsilon\\
\varepsilon & 1
\end{array}\right]\right)_{\left|g1\right\rangle ,\left|e0\right\rangle },h\left(n\nu_{O}I_{3}-\alpha\left[\begin{array}{ccc}
0 & \sqrt{n}\varepsilon & 0\\
\sqrt{n}\varepsilon & 1 & \sqrt{2\left(n-1\right)}\varepsilon\\
0 & \sqrt{2\left(n-1\right)}\varepsilon & 1
\end{array}\right]\right)_{Bn}\end{array},
\end{equation}

\end{widetext}with $I_{k}$ the identity matrix of dimension $k$
and $B_{n}=\left\{ \left|gn\right\rangle ,\left|e\left(n-1\right)\right\rangle ,\left|f\left(n-2\right)\right\rangle \right\} $
the basis for $n\geq2$. The diagonalization of each block yields
analytical eigenenergies and vectors, which for $n\geq2$ are functions
of the three real solutions of the cubic equation $x^{3}-2x^{2}+\left[1+2\epsilon^{2}-3n\epsilon^{2}\right]x+n\varepsilon^{2}=0$.
To shed light on the physics, we expand these analytical quantities
in the small parameter $\varepsilon$. As shown in Fig.~\ref{fig:resonant_ef},
the levels form three distinct energy ladders$\left\{ \left|\widetilde{gn}\right\rangle \right\} $,
$\left\{ \left|-n\right\rangle \right\} $ and $\left\{ \left|+n\right\rangle \right\} $:
The ladder$\left\{ \left|\widetilde{gn}\right\rangle \right\} _{n\geq0}$
corresponds to the almost unperturbed oscillator when the 3LS is left
in its ground state. With eigenenergies and eigenvectors\begin{widetext}
\begin{equation}
\begin{array}{l}
E(\left|\widetilde{gn}\right\rangle )=nh\left[\widetilde{\nu}_{O}+(n-1)K/2\right]+o(\epsilon^{5})\\
\left|\widetilde{gn}\right\rangle =\left[1-n\varepsilon^{2}/2,-\sqrt{n}\varepsilon,\sqrt{2n\left(n-1\right)}\varepsilon^{2}\right]_{Bn}+o(\epsilon^{3})
\end{array},\label{eq:effective oscillator}
\end{equation}
\end{widetext}this effective oscillator $\tilde{O}$ has a shifted
frequency $\widetilde{\nu}_{O}=\nu_{O}+\alpha\left(\varepsilon^{2}-\varepsilon^{4}\right)$,
and a small Kerr non-linearity $K=2\alpha\varepsilon^{4}$ inherited
from the 3LS. The two other ladders $\left\{ \left|\pm n\right\rangle \right\} $
(extended down to $n=0$ by $\left|+0\right\rangle \equiv\left|-0\right\rangle \equiv\left|\widetilde{g0}\right\rangle $
and $\left|+1\right\rangle \equiv\left|-1\right\rangle \equiv\left|\widetilde{e0}\right\rangle $)
have energies and eigenvectors \begin{widetext} 

\begin{align}
\begin{array}{l}
\{E(\left|\pm n\right\rangle )\}=\{0,h\left[\nu_{O}-\alpha\left(1+\varepsilon^{2}\right)+o(\varepsilon^{3})\right],...,h\left[n\nu_{O}-\alpha\left(1\pm\sqrt{2\left(n-1\right)}\varepsilon+n\varepsilon^{2}/2+o(\varepsilon^{3})\right)\right]\}\\
\{\left|\pm n\right\rangle \}=\left|g0\right\rangle ,\left[\begin{array}{c}
\varepsilon\\
1-\frac{\varepsilon^{2}}{2}
\end{array}\right],...,\left[\begin{array}{c}
\epsilon\sqrt{\frac{n}{2}}\pm\sqrt{\frac{n}{n-1}}\frac{7n-8}{8}\varepsilon^{2}\\
\frac{1}{\sqrt{2}}\pm\frac{n}{8\sqrt{n-1}}\varepsilon+\frac{\left(33n-32\right)n}{64\sqrt{2}\left(n-1\right)}\varepsilon^{2}\\
\pm\frac{1}{\sqrt{2}}-\frac{n}{8\sqrt{n-1}}\varepsilon+\frac{n^{2}}{64\sqrt{2}\left(n-1\right)}\varepsilon^{2}
\end{array}\right]_{Bn}+o(\varepsilon^{3})
\end{array}.
\end{align}
\end{widetext}For $n\geq2$ the zeroth-order approximation in $\epsilon$
of these eigenvectors are simply the symmetric and anti-symmetric
superposition of $\left|e\left(n-1\right)\right\rangle $ and $\left|f\left(n-2\right)\right\rangle $.
A particular ``Fock state'' $\left|\widetilde{gn}\right\rangle $
of $\tilde{O}$ is selectively manipulable by addressing the $\left|\widetilde{gn}\right\rangle \rightarrow\left|\pm\left(n+1\right)\right\rangle $
transitions to the hybridized oscillator-transmon states, at frequencies
$\nu_{\pm n}=\nu_{ge}\pm\sqrt{2n}g_{O}+(3n+1)g_{O}\varepsilon/2+o(\varepsilon^{2})$
(with a Rabi frequency about $\sqrt{2}$ slower for all $n>0$ than
for the pure $\left|g\right\rangle \rightarrow\left|e\right\rangle $
transition). Note that these frequencies $\nu_{\pm n}$ do not vary
linearly with $n$ as in the usual dispersive case, but as $\sqrt{n}$.
Selective addressing of $\left|\widetilde{gn}\right\rangle $ requires
the separation $\varDelta\nu_{\pm n}=\nu_{\pm\left(n+1\right)}-\nu_{\pm n}$
to be larger than the transition linewidth. A second condition is
that the driving strength of the $\nu_{\pm n}$ transition is low
enough to avoid driving off-resonantly the neighboring transitions
at $\nu_{\pm\left(n+1\right)}$. 

It is now interesting to compare the Kerr non-linearity $K=2\alpha\left(g_{O}/\Delta\right)^{4}$
obtained here with the value $K'=2\Delta^{'}\left(g_{O}^{'}/\Delta^{'}\right)^{4}$
that would be obtained for a two-level system or the value $K"=\alpha^{"}\left(g_{O}^{"}/\Delta^{"}\right)^{4}$
obtained in perturbation for a transmon in the far dispersive regime
$g_{O}^{"},\alpha^{"}\ll\varDelta^{"}$ \cite{Black box}, keeping
the same separation $S=\sqrt{2}g_{O}=2\alpha^{"}\left(g_{O}^{"}/\Delta^{"}\right)^{2}$
between the first two Fock state dependent excitation frequencies.
With respect to the far dispersive case, the new non-linearity is
thus reduced by a factor $K''/K=\left(\alpha/\alpha^{''}\right)\left(\alpha/S\right)^{2}/2$
that can be made large easily. This reduction factor, which reaches
several hundreds (at fixed transmon anharmonicity $\alpha=\alpha^{"}$)
in our implementation of section \ref{sec:Experimental-implementation},
is what makes our 'ef-resonant' scheme interesting. What we show here
with a simple ef-resonant 3-LS is that getting out of the perturbation
regime $\alpha^{"}\ll\varDelta^{"}$ reduces drastically the Kerr
non-linearity. However considering only three levels makes the argumentation
only qualitative for a transmon at large number of photons in the
oscillator, and a quantitative evaluation requires taking into account
at least the fourth transmon level as we do in section \ref{sec:Experimental-results}.

\section{Experimental implementation\label{sec:Experimental-implementation}}

\begin{figure}
\includegraphics[width=8cm]{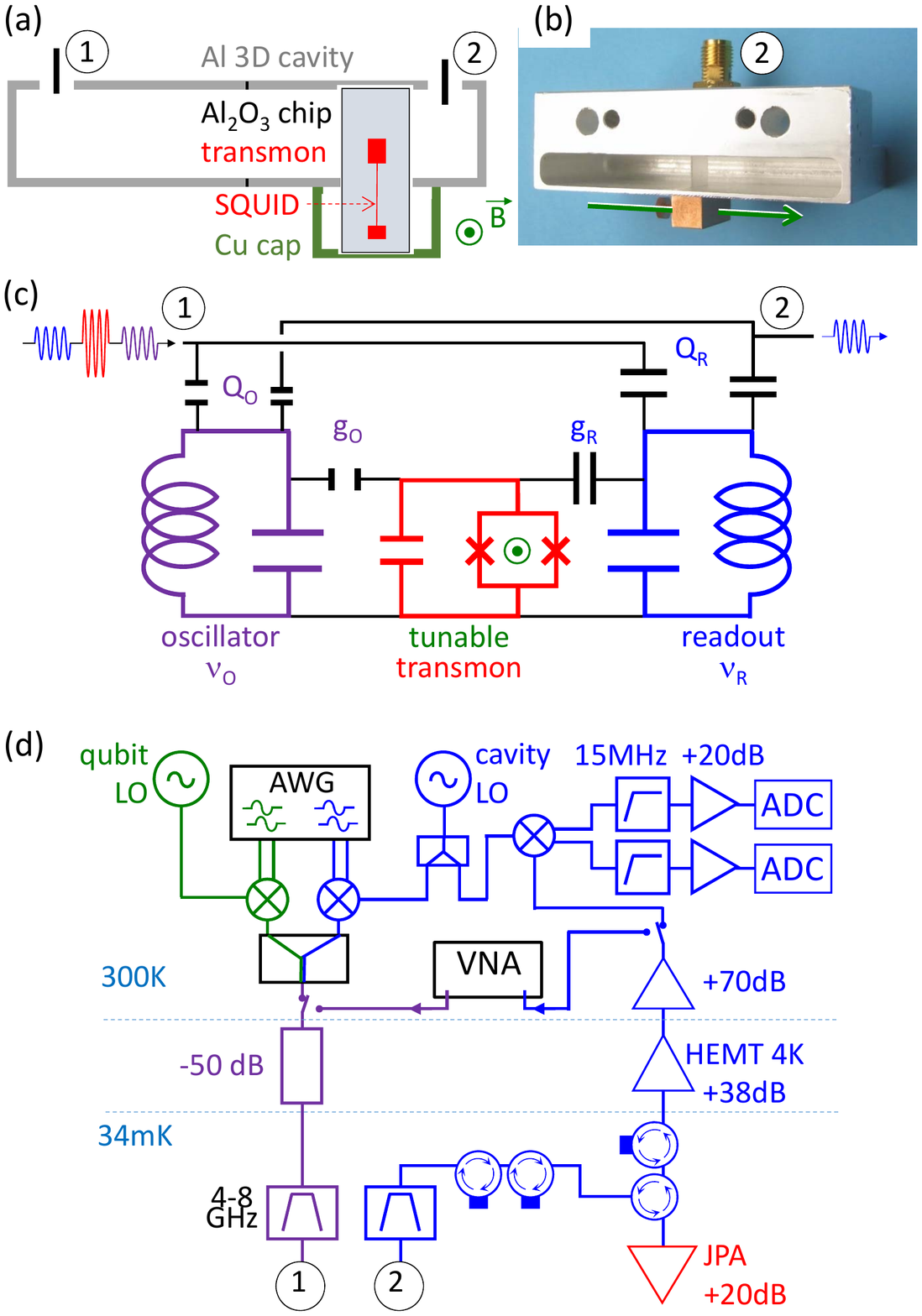} \protect\caption{\label{fig:setup}Circuit-QED implementation of the 'ef-resonant'
scheme. (a) The harmonic oscillator $O$ is the mode 120 of a superconducting
Al cavity and the MLS is a tunable transmon qubit with a SQUID. The
transmon chip is inserted only partly in the cavity, the SQUID being
exposed to a dc magnetic field $\protect\overrightarrow{B}$. The
transmon is weakly coupled to $O$ and strongly to the cavity mode
210 used for dispersive readout of the transmon state. The two modes
have quality factors $Q_{O}=2\times10^{6}$ and $Q_{R}=15\times10^{3}$.
(b): Picture of one half-cavity with chip and Cu cap. (c) Equivalent
electric circuit of the system with relevant frequencies $f_{R,O}$,
quality factors $Q_{R,O}$, and coupling frequencies $g_{R,O}$. $O$
is driven coherently and resonantly through port 1 (purple pulse).
Fock state $\left|gn\right\rangle $ are manipulated by driving the
$\left|\widetilde{gn}\right\rangle \rightarrow\left|+n\right\rangle $
transition (red pulse). A projective measurement on Fock state $\left|\widetilde{gn}\right\rangle $
is obtained by a $\pi$ pulse at $\nu_{+n}$ followed by a readout
pulse (in blue). (d) Electrical setup at room temperature (300K) and
inside the dilution refrigerator: Cavity resonances are measured with
continuous waves using a vectorial network analyzer (VNA) whereas
pulsed experiments use heterodyne modulation and demodulation. Microwave
pulses at the cavity (qubit) frequencies $\nu_{O,R}$ ($\nu_{+n}$)
are obtained by single sideband mixing of a continuous microwave (LO)
with an intermediate frequency modulated pulse generated by two channels
of an arbitrary waveform generator (AWG). All pulses travel along
an attenuated and filtered line to cavity port 1. The readout signal
transmitted at port 2 is filtered, isolated from backward propagating
noise, amplified with a parametric amplifier (JPA) in reflection,
a high electron mobility transistor (HEMT), and room temperature amplifiers,
then demodulated to produce two quadratures, which are finally filtered,
amplified and digitized (ADC).}
\end{figure}

We implement the proposed 'ef-resonant' scheme in a three-dimensional
circuit-QED setup \cite{Paik cavity 3D} combining a cavity with input
(1) and output (2) ports and a tunable transmon qubit \cite{Koch-Transmon}
(see Fig.~\ref{fig:setup}a-b). To be superconducting at low temperature
and have a high internal quality factor, the cavity is made of two
blocks of pure aluminum, which are milled, pierced, polished, and
chemically etched over about 20\,\textmu m. The transmon is fabricated
on sapphire by double-angle evaporation of Al and oxidation, through
a suspended shadow mask made by e-beam lithography. It has two pads
connected by a 2.6\,mm long wire including a magnetic flux tunable
Josephson junction with a SQUID geometry located 50\,\textmu m above
the bottom pad. This enables tuning the transmon energy spectrum and
reaching the 'ef resonant' condition. The transmon is only partly
inserted in the cavity so that the SQUID remains about 0.1\,mm outside,
in the applied external magnetic field. More precisely, the bottom
part of the transmon with the SQUID is held and protected by a copper
block, the other side being inserted in the cavity through a slit
in the bottom wall. The two halves of the cavity are then pressed
one against the other with an indium seal in-between.

In our design the TE120 cavity mode is used as the quantum oscillator
$O$ at frequency $\nu_{O}\sim6.4\thinspace\mathrm{GHz}$ whereas
mode TE210 at frequency $\nu_{R}\sim7.3\thinspace\mathrm{GHz}$ is
used for reading the quantum state of the transmon dispersively \cite{Blais circuit QED,Wallraff CircuitQED}.
The transmon and the ports are thus placed very close to a node of
mode $O$ so that the corresponding coupled quality factor $Q_{O}$
is dominated by the internal losses, and the transmon-oscillator coupling
$g_{O}$ is about 10\,MHz. They are also placed at an antinode of
mode $R$ to get a strong transmon-readout coupling $g_{R}\sim150\thinspace\mathrm{MHz}$
and thus a large enough dispersive shift $\chi_{R}\sim10\thinspace\mathrm{MHz}$,
and a low coupled quality factor $Q_{R}\sim15\times10^{3}$ allowing
fast readout of the transmon. The exact positions as well as the precise
transmon geometry are determined using the CST 3D microwave simulator
and methods adapted from \cite{Black box}. Attention is paid to keep
the transmon's charging energy (one electron) below $300\thinspace\mathrm{MHz\times h}$
in order to avoid variations of the level $f$ energy due to charge
parity fluctuations \cite{Riste - Parity}. For better measurement
efficiency, output port 2 is about 6 times more strongly coupled to
each mode than input port 1.

The equivalent circuit of the system is shown in Fig.~\ref{fig:setup}c.
The transmon-cavity system is mounted inside a coil placed in a mu-metal
shield, and is attached to the cold plate of a cryofree dilution refrigerator
with base temperature 35\,mK. It is connected to the electrical setup
of Fig.~\ref{fig:setup}d, which includes a home made quantum limited
Josephson parametric amplifier (JPA) similar to \cite{Zhou JPA}.
Simple continuous microwave measurements at a single frequency are
done with a vectorial network analyzer (VNA), whereas pulsed measurements
involving $\nu_{O}$, $\nu_{R}$, and one or two transmon frequencies
use heterodyne modulation and homodyne demodulation as described in
Fig.~\ref{fig:setup}d.

\section{Experimental results\label{sec:Experimental-results}}

\subsection{Spectroscopic characterization}

\begin{figure*}
\includegraphics[width=15cm]{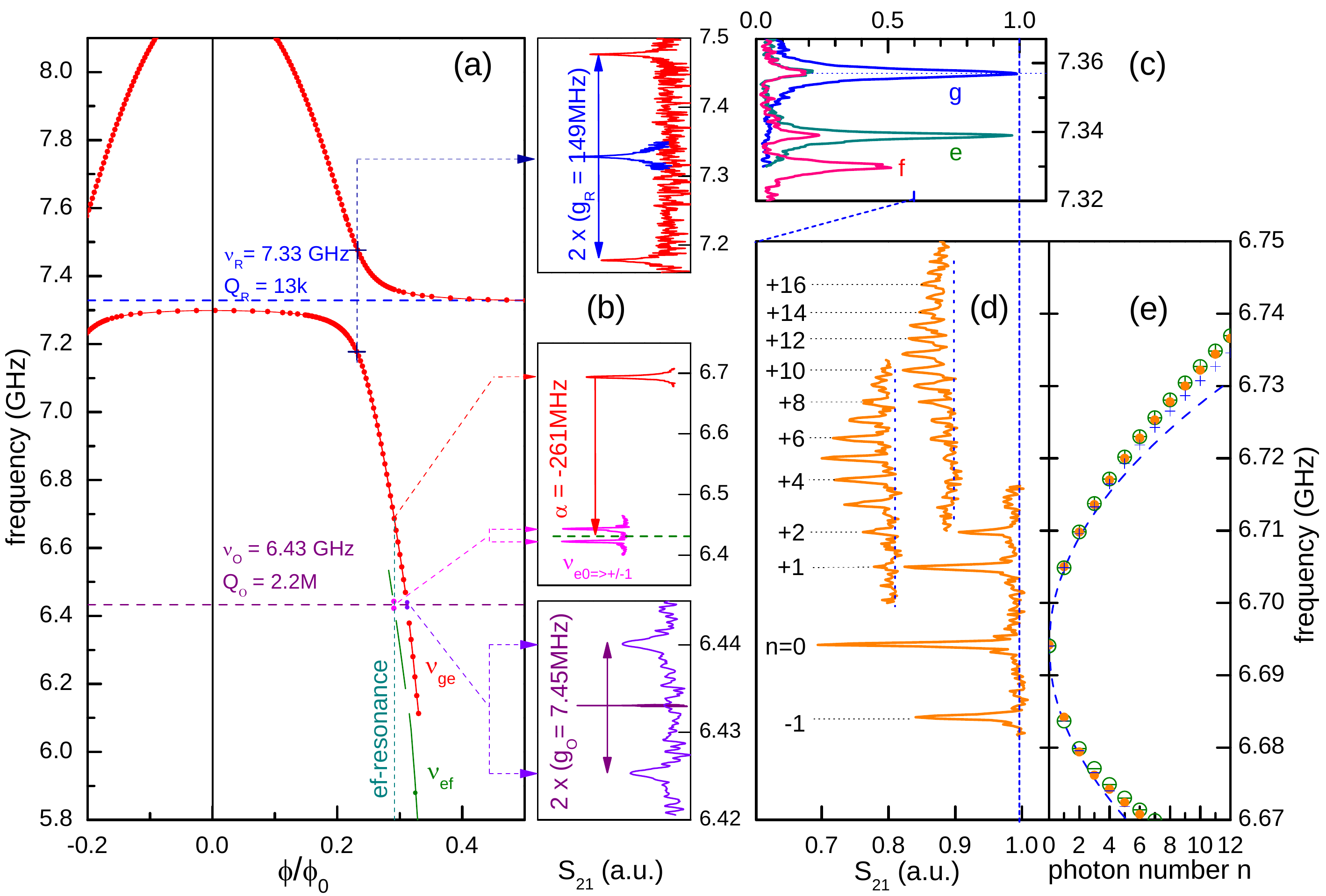} \protect\caption{\label{fig:spectro}Spectroscopic characterization of the system.
(a) Measured transition frequencies as a function of the magnetic
flux $\phi$ applied through the transmon SQUID. Blue and purple horizontal
dashed lines correspond to the readout and oscillator frequencies
$\nu_{R}$ and $\nu_{O}$, red dots to the hybridization of $\nu_{ge}$
and $\nu_{R}$, and the green line to $\nu_{ef}$. The 'ef-resonant'
condition is obtained at $\nu_{O}=\nu_{ef}$. (b) Spectra determining
the qubit-resonator coupling frequencies $g_{R,O}$, as well as the
qubit anharmonicity $\alpha$. Top: spectra around $\nu_{R}$ with
$\nu_{ge}$ either far away (blue peak) or anti-crossing $\nu_{R}$
(two red peaks). Bottom: spectra around $\nu_{O}$ with $\nu_{ge}$
either far away (purple peak) or anti-crossing $\nu_{O}$ (two violet
peaks). Middle: spectrum at the 'ef-resonant' working point showing
the $ge$ (red) and $\left|\widetilde{e0}\right\rangle \rightarrow\left|\pm1\right\rangle $
(magenta) transitions. (c) Readout resonance when the qubit is left
in $\left|g\right\rangle $ or excited in $\left|e\right\rangle $
or $\left|f\right\rangle $. Note that the residual thermal population
of level e is well below 1\%. The dashed line indicates the frequency
at which the readout mode transmission $S_{21}$ is measured in panel
(d). (d) Qubit spectra measured at the 'ef-resonant' point for three
different fillings of $\tilde{O}$ ($\beta\simeq0.54$, 4.5 and 10
- last two horizontally shifted for clarity), showing transitions
$\left|\widetilde{gn}\right\rangle \leftrightarrow\left|\pm n\right\rangle $
(noted $\pm n$) from $-n=-1$ to $n=17$. (e) Transition frequencies
$\nu_{\pm n}$ deduced (orange dots) from spectroscopy (d), calculated
in section \ref{sec:Theoretical-summary} (+ and - symbols), and numerically
computed by diagonalization of the system Hamiltonian (open circles).
Lines correspond to the parabolic approximation $\nu_{\pm n}\simeq\nu_{ge}\pm\sqrt{2n}g_{O}$.}
\end{figure*}

Once at 35\,mK, the system is first characterized with the VNA as
a function of the current in the coil (see Fig. \ref{fig:spectro}a-b).
The lowest transition frequencies of the hybridized readout mode-transmon
system yields two peaks at frequencies (red points) that are periodic
in flux $\Phi$, one period corresponding to one flux quantum $\Phi_{0}=h/2e$.
Away from the avoided crossing, these peaks tend to $\nu_{R}$ and
$\nu_{ge}$. At $\Phi\simeq0.5\Phi_{0},$ modes $R$ and $O$ are
unperturbed and one gets their frequencies (central peaks of panels
b-top and b-bottom) as well as the total quality factor $Q_{O}=2.2\times10^{6}$
of the oscillator, which corresponds to an energy relaxation time
$T_{O}=54\thinspace\lyxmathsym{\textmu}s$. At $\Phi=0.23\Phi_{0}$
and $0.31\Phi_{0}$ the transmon frequency $\nu_{ge}$ anti-crosses
$\nu_{R}$ and $\nu_{O}$, yielding the double peaks of panels b-top
and b-bottom, separated by twice the coupling frequencies $g_{R}=149\thinspace\mathrm{MHz}$
and $g_{O}=7.45\thinspace\mathrm{MHz}$. The setup for pulsed spectroscopy
is then used to excite the transmon-oscillator system with one or
several pulses, and then detect this excitation from a change of the
transmission $S_{21}(\nu_{R})$ of the readout mode (see pulses in
Fig. \ref{fig:setup}c ). Figure \ref{fig:spectro}c shows how the
readout line is dispersively shifted when the transmon is left in
$\left|g\right\rangle $, partly excited in $\left|e\right\rangle $
with a single pulse at $\nu_{ge}$, or partly excited in $\left|f\right\rangle $
with the same first pulse and a second one at $\nu_{ef}$. The different
readout lines g,e and f do not overlap so that at the top of any peak,
a change of the qubit state induces a complete suppression of the
transmission. The $\nu_{ef}\left(\Phi\right)$ dependence (green line
in panel a) is thus obtained by finding first $\nu_{ge}\left(\Phi\right)$
and then scanning for each $\Phi$ a second tone around $\nu_{ef}$
and measuring $S_{21}(\nu_{R,e})$ at the top of the e readout peak.

The 'ef-resonant' condition that we target occurs at $\Phi_{ef-r}=0.291\Phi_{0}$
(vertical cyan line in panel a), when the $ef$ transition crosses
$\nu_{O}$, leading to the hybridization discussed in section \ref{sec:Theoretical-summary}
and to the splitting of the $ef$ peak into two symmetric peaks $\nu_{e0\rightarrow\pm1}$
(shown in magenta in panel b-middle). The dataset of panel b yields
the effective anharmonicity $\alpha=-261\thinspace\mathrm{MHz}$ ('effective'
means here in presence of the additional readout mode that shifts
dispersively the transmon levels). From now on, the system is further
characterized at $\Phi_{ef-r}$ with $S_{21}(\nu_{R,g})$ being measured
at the top of the $g$ readout peak (see panel c). At this point,
we obtain the qubit relaxation time $T_{1}=1.9\thinspace\mathrm{\mu s}$
and coherence time $T_{2}^{*}=2.4\thinspace\mathrm{\mu s}$. The photon
resolved transition frequencies $\nu_{\pm n}$ are then found by filling
mode $\widetilde{O}$ with a coherent state $\left|\beta\right\rangle =\sum\sqrt{p_{\beta}\left(n\right)}\left|n\right\rangle $
using a first resonant excitation pulse at $\nu_{\widetilde{O}}$
and scanning the frequency of a second pulse about $\nu_{ge}$. Panel
d shows the peaks at frequencies $\nu_{n}$ for $n=-1,0,1,...,16$,
resulting from three different values of $\beta$; the peak amplitudes
for each $\beta$ approximately reproduce the Poisson distributions
$p_{\beta}\left(n\right)$ expected for coherent states. As opposed
to the dispersive case and as expected, the peaks get nearer to each
other with increasing n: in panel e their frequencies (orange dots)
are compared to the analytical expression of section \ref{sec:Theoretical-summary}
(calculated with three transmon levels) and to the $\nu_{\pm n}$
values resulting from the numerical diagonalization of an effective
oscillator-transmon Hamiltonian also including the fourth transmon
level $h$ (the numerical diagonalization with only three levels coincide
with analytical results). A good agreement is found between the experiment
and the effective four-level transmon model using the measured values
of $\nu_{O}$, $\nu_{ge}(\Phi_{ef-r})$, $\alpha$ and $g_{O}$, as
well as the shifted energy $h\left(3\nu_{ge}-848\thinspace\mathrm{MHz}\right)$
of eigenstate $\left|h\right\rangle $ calculated from $\nu_{R}(\Phi_{ef-r})$
and $g_{R}$.

\subsection{Oscillator field characterization by selective $\pi$ pulses on $+n$
transitions}

\begin{figure}[!t]
\includegraphics[width=7cm]{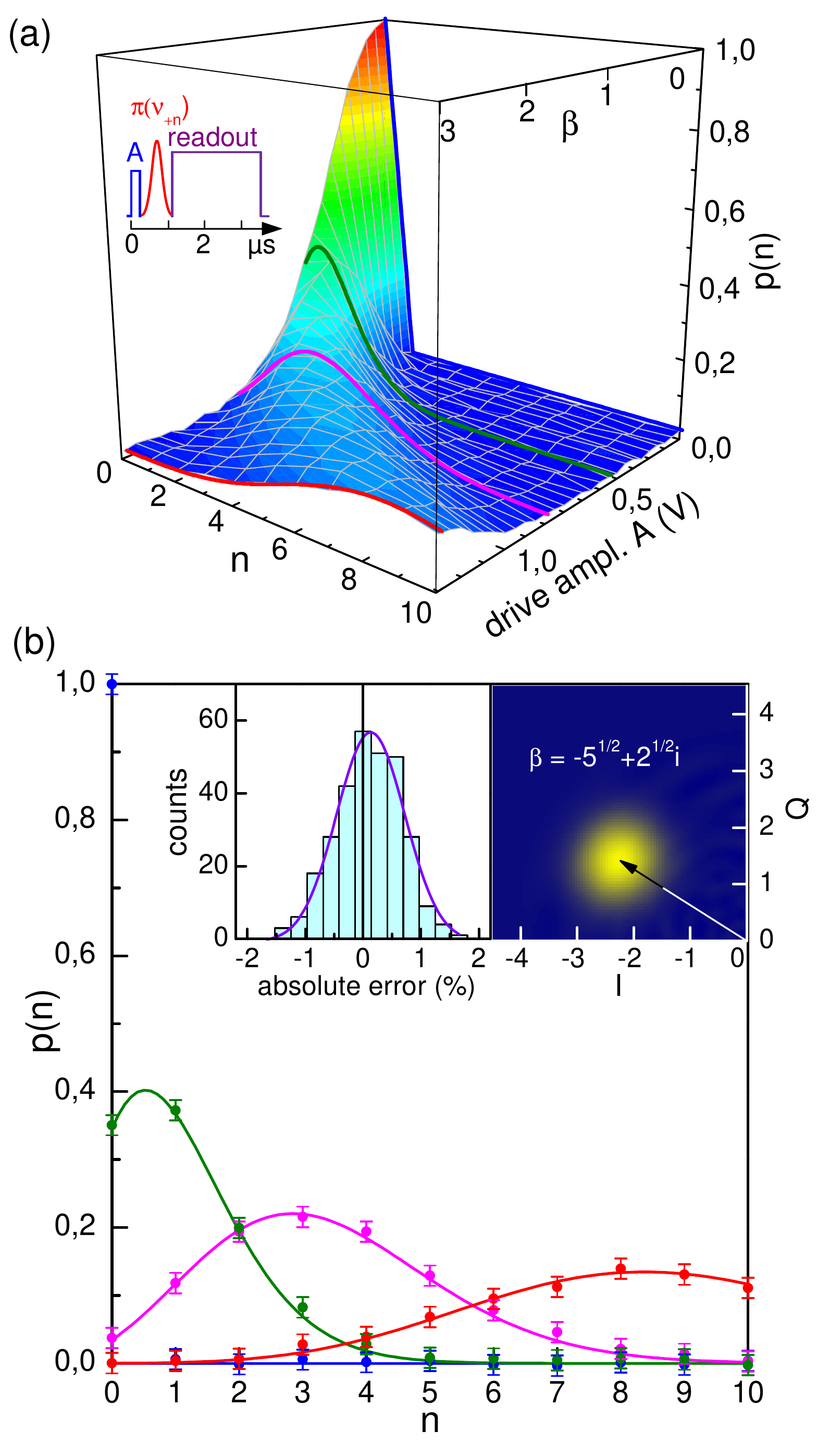} \protect\caption{\label{fig:coherent}Generation and measurement of coherent states
$\left|\beta\right\rangle $ in the oscillator mode: a state is obtained
from a 0.3\,\textmu s long rectangular pulse with frequency $\nu_{\widetilde{O}}$
and amplitude $A$ (expressed here in Volt on the AWG of fig. \ref{fig:setup}d);
a $\pi$ pulse on one of the $+n$ transition then excites conditionally
the transmon and the variation $\Delta S_{21}(\nu_{R},A,n)$ of the
cavity transmission is measured. (a) Occupation probability $p(n)$
of Fock state $\left|n\right\rangle $ for $n\in\left[0,10\right]$
and for increasing $A$. $p(n)$ is obtained by dividing $\Delta S_{21}(\nu_{R},A,n)$
by a calibration factor $c_{n}$ resulting from a fit of the coherent
model $\beta(A)=kA$ to the whole dataset: the fitted coefficients$\{c_{0},...,c_{10}\}=\{$0.764,
0.835, 0.847, 0.846, 0.833, 0.854, 0.846, 0.834, 0.847, 0.832, 0.841$\}$
differ by less than 3\%, except for $c_{0}$ that corresponds to a
transition between transmon states not hybridized with mode $O$.
(b) $p(n)$ cuts (also shown in a) at $A=0$ (blue), 0.45 (green),
0.8 (magenta) and 1.3\,V (red) showing both measured data (dots with
$\pm2\sigma$ error bars) and expected Poisson distributions (lines).
Residual errors between data and fit are homogeneously distributed
all over the dataset (not shown) and Gaussianly distributed with a
standard deviation $\sigma=0.6\%$ and a shift of 0.1\% (left inset).
Right inset is the reconstruction of the Wigner function of a targeted
state $\left|\beta=-\sqrt{5}+i\sqrt{2}\right\rangle $ by tomography
and maximum likelihood analysis (see text).}
\end{figure}

Having characterized the system energy diagram, we now address individually
the photon number resolved transitions $+n$ to fully characterize
the field and probe the harmonic character of oscillator $\widetilde{O}$:
as an example, we fill $\widetilde{O}$ with coherent states $\left|\beta\right\rangle $
and retrieve them by standard quantum state tomography. This tomography
involves the measurement of the occupation probabilities $p(n)$ of
several Fock states $\left|n\right\rangle $. As done in \cite{Kirchmair-TomoP(n)}
for the dispersive case, the Fock state population $p(n)$ is simply
transferred to a qubit excited state, which is then read out. In our
'ef-resonant' scheme this transfer between $\left|\widetilde{gn}\right\rangle $
and $\left|+n\right\rangle $ consists in applying a $\pi$ pulse
on the $+n$ transition. These $\pi$ pulses at frequencies $\nu_{+n}$
have a Gaussian shape with $140\thinspace\mathrm{ns}$-long standard
deviation, and are calibrated in amplitude for $n\in\left[0,10\right]$.

Because of many possible imperfections in the transfer or readout
process of $p\left(n\right)$, such as different relaxation times
of $\left|+n\right\rangle $ for different $n$ during drive and/or
readout, the $p\left(n\right)$ measurement method is carefully calibrated
as described now: A series of rectangular pulses with fixed frequency
$\nu_{\widetilde{O}}$, fixed 0.3\,\textmu s long duration, and increasing
amplitudes $A$ are used to generate a priory coherent states $\left|\beta\right\rangle $
in $\widetilde{O}$; immediately after, a $\pi$ pulse is applied
on one of the $+n$ transition, and the relative decrease $s(n,A)=\triangle S_{21}(\nu_{R,g})$
of the readout mode transmission is measured with a final pulse; this
signal, averaged over 40000 sequences (separated by $\sim6T_{O}$
to let the field relax to its ground state) is measured for $n\in\left[0,10\right]$
and for 26 values of $A$. Then a model assuming that the coherent
amplitude $\beta=kA$ is proportional to the input amplitude $A$
and that the raw signal $s(n,A)$ reproduces the Poisson distribution
$p_{\beta}\left(n\right)$ of coherent states $\left|\beta\right\rangle $
up to calibration coefficients $c_{n}$ that depend only on $n$,
is fitted to the whole data set. The fit $s(n,A)=c_{n}p_{kA}\left(n\right)$,
shown in Fig. \ref{fig:coherent}, yields the eleven parameters $\{c_{0},...,c_{10}\}$
as well as the filling rate $k=2.29\thinspace\mathrm{V}^{-1}$. The
residual error of the fit is homogeneously and Gaussianly distributed
with a standard deviation of only 0.6\%, which confirms the validity
of the model.

The calibration coefficients $c_{n}$ being known, the occupation
probabilities $p_{\rho}\left(n\right)=s(n,\rho)/c_{n}$ can now be
measured to fully characterize any state $\rho$ of the oscillator
field using standard quantum field tomography \cite{QSTomo} and maximum
likelihood techniques \cite{Hradil}. As a demonstration, we target
a coherent state $\left|\beta=-\sqrt{5}+i\sqrt{2}\right\rangle $,
prepare it using a coherent rectangular pulse with proper amplitude
and phase, and then measure it. This is done by recording the Fock
state probabilities $p(n,\gamma)$ for $n\in\left[0,7\right]$ and
for 240 different complex displacements $\gamma$ of $\left|\beta\right\rangle $,
and then reconstructing the field density matrix $\rho$ in a Hilbert
space truncated to 18 photons, by maximizing the likelihood of the
$\left\{ p(n,\gamma)\right\} $ dataset. The corresponding Wigner
function is shown in the bottom inset of Fig.~\ref{fig:coherent}b.
The fidelity $\mathrm{Tr}\left(\sqrt{\sqrt{\rho}\left|\beta\right\rangle \left\langle \beta\right|\sqrt{\rho}}\right)$
of the reconstructed $\rho$ to the targeted state is of order 98\%
immediately after the calibration (the calibration has to be done
every three days, typically).

\subsection{Non-linearities of the oscillator}

\begin{figure*}
\includegraphics[width=13cm]{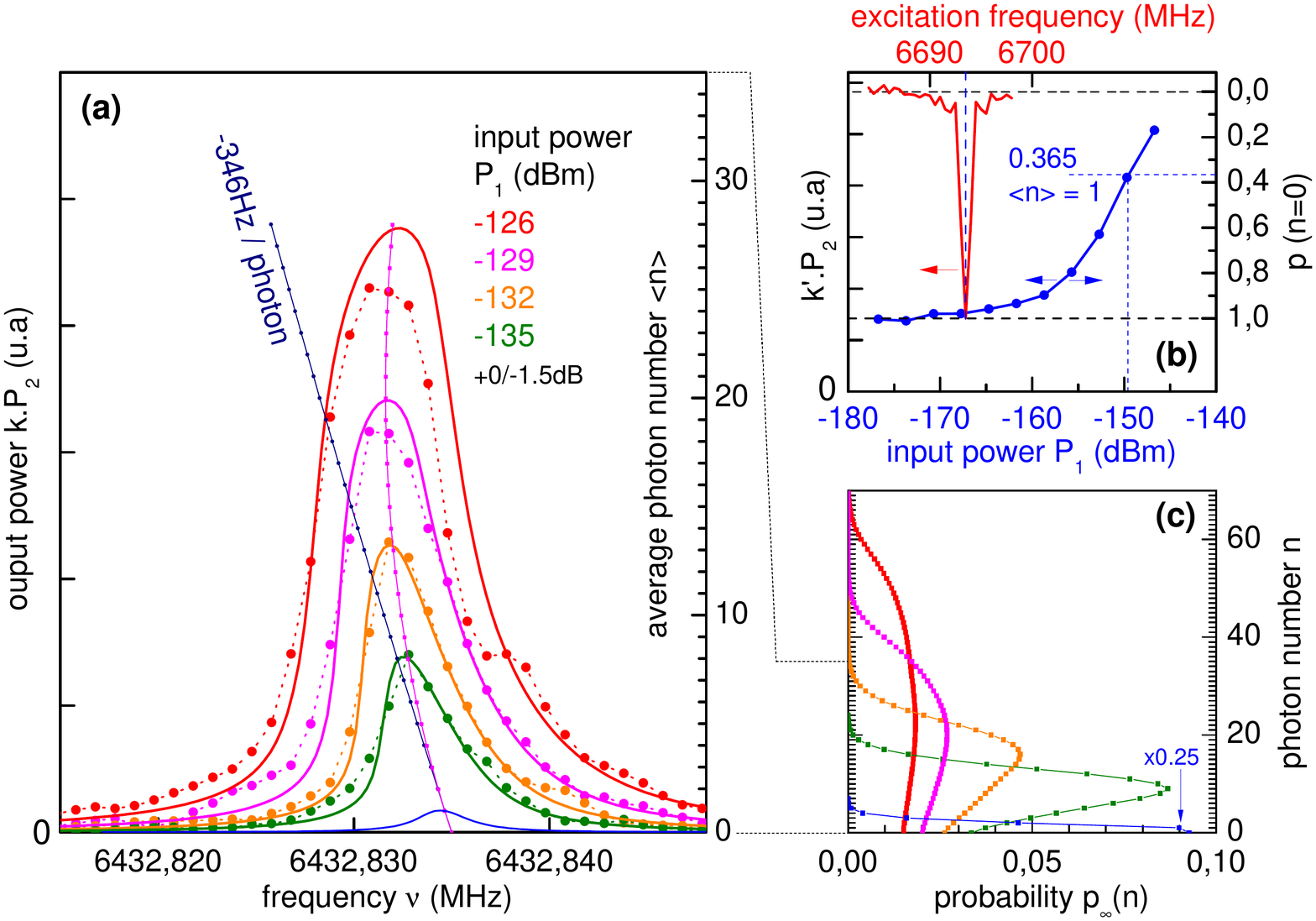} \protect\caption{\label{fig:linearity}Steady state non-linearities of oscillator $\widetilde{O}$
in the 'ef-resonant' condition. (a) Transmitted power (in linear units)
measured with the VNA (dots joined by dotted lines) for input powers
$P_{1}$ varying from -135\,dBm to -126\,dBm by steps of 3dB. Solid
lines are the steady state average photon number $<n>$ in mode $\widetilde{O}$
simulated numerically (see text) for the same $P_{1}$ values according
to calibration, and for $P_{1}=-149.6\thinspace\mathrm{dBm}$ ($<n>=1$)
at resonance. The non-linearity in frequency calculated with four
transmon levels (magenta line) is smaller than -346\,Hz/photon Kerr
constant (dark blue line). A large non-linearity in power is however
observed (see text). b) Calibration of $P_{1}$ versus $<n>$. Red:
variation of the transmitted readout amplitude $\left|P_{2}\left(\nu\right)\right|$
when keeping mode $\widetilde{O}$ in its ground state and detuning
the qubit excitation frequency $\nu$ of a $\pi$ pulse away from
the $+0$ transition at $\nu_{ge}$ (dashed vertical line). Blue:
Same readout amplitude after the resonant qubit $\pi$ pulse when
filling $\widetilde{O}$ by a continuous tone with frequency $\nu_{\tilde{O}}$
and variable input power $P_{1}$. The average photon number $<n>$
reaches 1 at $P_{1}=-149.6\thinspace\mathrm{dBm}$ (blue dashed lines
- see text). (c) Simulated photon number distributions $p_{\infty}(n)$
at the top of each simulated resonance of panel (a). The blue curve
for $<n>=1$ has been multiplied by 0.25 for clarity.}
\end{figure*}

We now check experimentally our claim of a very small Kerr effect
for the 'ef-resonant' scheme. With the $g_{O}=7.45\thinspace\mathrm{MHz}$
and $\alpha=-261\thinspace\mathrm{MHz}$ values determined spectroscopically,
the Kerr non-linearity calculated for the three level transmon model
of section \ref{sec:Theoretical-summary} is $K=-346\thinspace\mathrm{Hz}$
per photon. However, when diagonalizing numerically the Hamiltonian
of the system including the fourth transmon energy level, the Kerr
effect is found to be even smaller, to depend on $n$, and to cancel
and reverse its sign at about 20 photons. Such an ultra-small Kerr
effect would yield no sizable phase accumulation of the different
Fock states over the cavity relaxation time $T_{O}=54\thinspace\lyxmathsym{\textmu}s$.
This makes it difficult to measure it dynamically, by recording either
the trajectory of a field state in phase space as in \cite{Yin - dynamic kerr,Kirchmair-TomoP(n)},
or the power dependence of the resonance line shape at short time
as in the supplementary information of \cite{Kirchmair-TomoP(n)}.
Consequently, we simply measure the steady-state transmitted power
$P_{2}$$\left(\nu\right)$ of mode $\widetilde{O}$ as a function
of the excitation frequency $\nu$ at several input powers $P_{1}$,
using the VNA with a narrow enough measuring bandwidth of $1\thinspace\mathrm{kHz}$.
The corresponding curves are shown in Fig.~\ref{fig:linearity}a
(dots) in arbitrary units of the output power $P_{2}$ (left scale).

Analyzing quantitatively the dataset requires a precise knowledge
of the average photon number $<n>$ in the resonator as a function
of $\nu$ and $P_{1}$. In this aim, we perform the following additional
in-situ calibration and data analysis. We first use the transmon to
determine experimentally the input power $P_{1,1}=-149.6\thinspace\mathrm{dBm}$
that populates the cavity with $<n>=1$, which corresponds to $p_{\beta=1}(n=0)=0.365$
(see Fig. \ref{fig:linearity}b) assuming a coherent steady state
$\rho_{1}=\left|\beta=1\right\rangle \left\langle \beta=1\right|$.
Then quantum simulations of the oscillator $O$ coupled to the effective
4 level transmon are performed using the QuTiP Python toolbox \cite{qutip},
the already mentioned measured parameters, and the calibrated $P_{1}$.
Solid lines in Fig.~\ref{fig:linearity}a show the resonance lines
$<n>(\nu)$ obtained with the steady state solver of QuTiP. By scaling
vertically the experimental curves so that simulation and experiment
match for $P_{1}=-135\thinspace\mathrm{dBm}$, we obtain a fair agreement
for all curves, leading to the following results: First the resonance
lines display indeed a very small Kerr effect, with a shift towards
lower frequency with increasing $P_{1}$ significantly smaller than
$K$ (oblique dark blue line), and changing sign between $<n>=15$
and 25, in good agreement with the calculated shift (magenta line).

A second and unanticipated effect is that although the non-linearity
in frequency is small, a large non-linearity in input power is observed,
with $P_{2}$ and $<n>$ increasing by a factor of only 3 when $P_{1}$
is increased by a factor 8 (see extreme curves in Fig.~\ref{fig:linearity}a).
Simulating the time evolution of the system initialized in its ground
state and driven coherently at $\nu_{\tilde{O}}$ reveals the cause
of this non-linearity in $P_{1}$: the small hybridization of the
oscillator with the transmon (see Eq. \ref{eq:effective oscillator})
that has finite coherence times $(T_{1},T_{2}^{*})$, progressively
induces Fock state dephasing. The increasing field perfectly coherent
at the beginning of the dynamics slowly becomes incoherent when approaching
the steady state,which reduces its amplitude. This is illustrated
in Fig.~\ref{fig:linearity}c by the photon number distributions
$p_{P1,t=\infty}\left(n\right)$ obtained from the steady state solver
at the top of each resonance curve of panel a: although $p_{P1,t=\infty}\left(n\right)$
corresponds almost exactly to the Poisson distribution for $\rho_{1}$
(which validates the calibration of $P_{1}$ in the previous section),
the other distributions for larger $<n>$ are less and less Poissonian.
Large coherent states can nevertheless be obtained at times shorter
than a few tens of $T_{2}^{*}$, as observed in the previous section
for a duration of the coherent drive $t\simeq T_{2}^{*}/8$. Note
that the transmon-induced cavity non-linearity in power observed and
simulated here also exists in the dispersive regime, and is an effect
that would deserve a theoretical evaluation.

\section{Conclusion}

We have described a way to manipulate the quantum state of a harmonic
oscillator by coupling it to an anharmonic multilevel system (MLS),
without paying the price of a large Kerr non-linearity of the oscillator
inherited from the MLS. We have demonstrated our 'ef-resonant' scheme
using a 3D circuit-QED setup, in a new geometry involving a tunable
transmon qubit partially inserted inside a single multimode superconducting
cavity. Fock state manipulation was demonstrated by quantum state
tomography of a coherent field in the cavity. The non-linearity was
measured to be very small, provided the total field manipulation time
is not much longer than the qubit coherence time. Our setup and coupling
scheme provide a new platform for manipulating at will mesoscopic
quantum fields inside a harmonic resonator, and producing non-classical
states in various ways. In particular, the ef-resonant scheme would
reduce the Kerr non-linearity of the promising platform proposed and
developed \cite{Vlastakis schrodinger cat 100,Leghtas Encoding,Wang entanglement 2 cavities}
for encoding quantum information in Schrödinger cat states of the
cavity field. We plan to use this scheme for demonstrating the quantum
Zeno dynamics of the cavity field as proposed in \cite{Raimond Zeno}.

\section*{Acknowledgment}

We gratefully acknowledge discussions within the Quantronics group,
technical support from P. Orfila, P. Senat, J.C. Tack, D. Duet, and
V. Padilla, as well as financial support from the European research
contracts CCQED and ScaleQIT.

\end{document}